# Data Partitioning View of Mining Big Data

Shichao Zhang

Guangxi Key Lab of Multi-source Information Mining & Security
College of Computer Science and Information Technology
Guangxi Normal University, Guilin, 541004, PR China
zhangsc@mailbox.gxnu.edu.cn

**Abstract**: There are two main approximations of mining big data in memory. One is to partition a big dataset to several subsets, so as to mine each subset in memory. By this way, global patterns can be obtained by synthesizing all local patterns discovered from these subsets. Another is the statistical sampling method. This indicates that data partitioning should be an important strategy for mining big data. This paper recalls our work on mining big data with a data partitioning and shows some interesting findings among the local patterns discovered from subsets of a dataset.
**Key Words:** Big data; big data mining; data partitioning; statistical sampling

## 1. Introduction

Big data has become a hot research area after the "Nature", one of top-end journals, published a special issue on big data, named "Science in the petabyte (PB) era" [Nature 2008]. Before this came up, there were many words/terms of standing for "big" arose in papers and articles, such as large (scale), huge, very large, massive and vast. Seems "big" is an easy name to be defined, accepted, understood and propagated. These reports also indicated that big data mining has widely been studied after data mining was proposed as a research field.

In fact, I had a firsthand experience of how it is actually difficult to in-memory identify frequent patterns in a large scale dataset, when I worked in the National University of Singapore in 1998. To attack this issue, we proposed a solution of mining large scale dataset based on data partitioning [Zhang and Wu 2001]. The main idea is sketched in Figure 1.

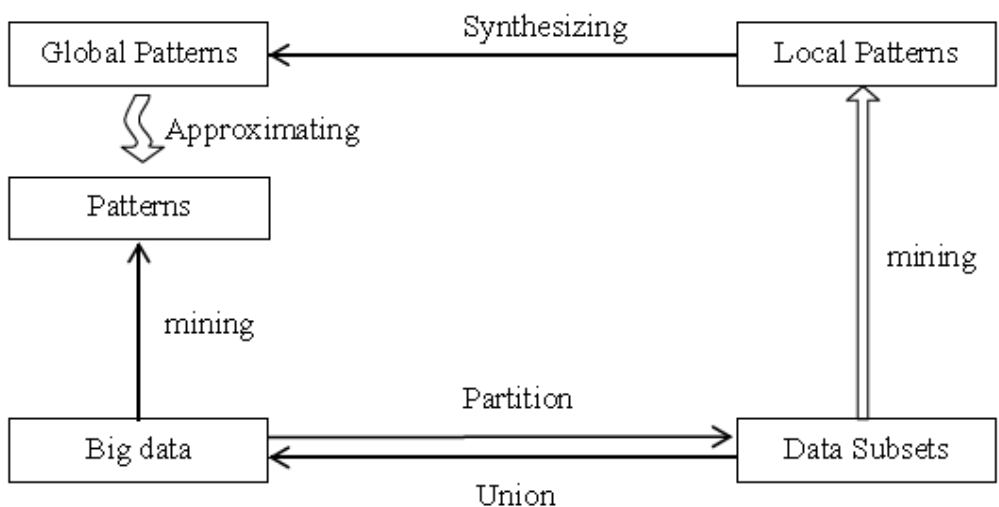

**Figure 1**. Mining big dataset based on data partitioning

To identify interesting patterns from a big dataset, our approximation is as follows. The big dataset is first partitioned to several subsets taking into account the memory size of computers. And then, each subset is mined in memory. These discovered patterns are referred to local patterns. Finally, all local patterns are fused to generate global patterns. These global patterns are output as the results of mining the big dataset. In practice, this is only an approximate solution of mining the big dataset, although data partitioning actually brings us an efficient approach. It is a lately research result that mathematically proved it is feasible to discover big data based on data partitioning [Xu, Zhang, Li 2015]. Therefore, I think data partitioning should be an important research direction of mining big data. In particular, data partitioning brings us some interesting findings among the local patterns discovered from subsets of a dataset, which cannot be learnt with traditional centralized-style mining methods.

## 2. What happen after mining segments of big data

After splitting a big dataset and mining its subsets segment by segment, there are many interesting patterns hided in these subsets, referred to local patterns in this paper. Those local patterns occurred in many segments can be synthesized as global patterns, i.e., "Pattern A" in Figure 2, where minsupport = 0.5 for all data subsets. They are really close to those frequent patterns that are directly discovered from the big dataset. However, most local patterns cannot be discovered from the big dataset with traditional centralized-style data mining methods. Some local patterns are often with high supports identified in few segments, like "Pattern B" in Figure 2. They should be called, such as subspace patterns in general, exceptional patterns for outlier detection, and burst pattern for mining historical big data. And some patterns look like trend patterns for dynamic data mining, i.e., "Pattern C" in Figure 2.

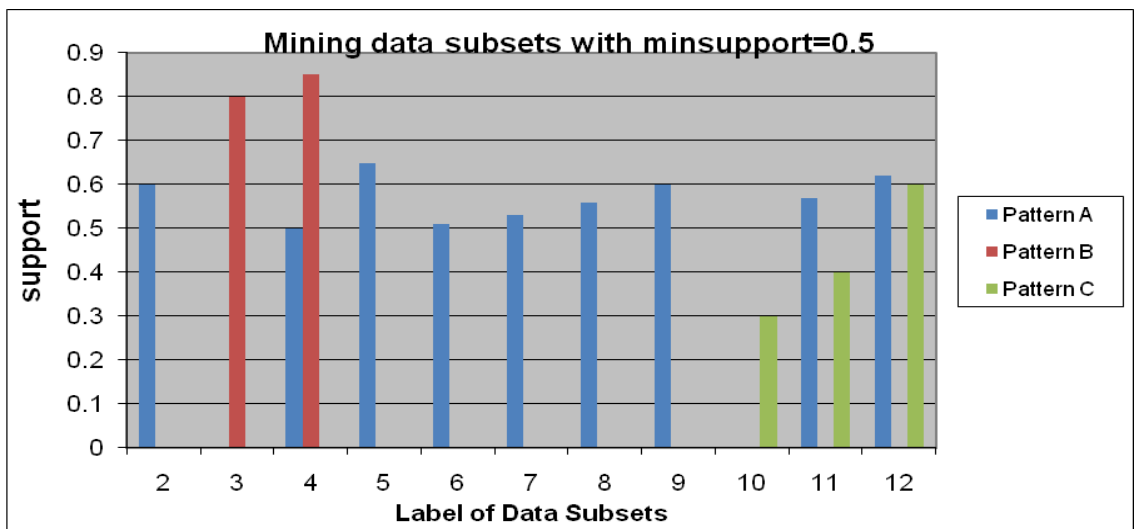

**Figure 2.** Patterns hided in data subsets of a big dataset

The above observations have showed what happen after mining segments of big data. There are really some interesting local patterns hidden in segments that cannot be discovered with traditional centralized-style mining methods. Do these interesting local patterns make sense in applications? In about 2003, I introduced my findings of mining big data to a manager of stock data processing in the UTS, Australia. The manager told me, these interesting local patterns are much more significant than traditional frequent patterns in real applications because it is difficult to obtain and master these interesting local patterns to the people in industrial community.

With these findings in our data partitioning view, we started some significant researches on mining big data. Main studies include approximate frequent patterns in [Zhang, Zhang and Webb 2003; Zhang and Zhang 2002], dynamic data mining in [Zhang, Zhang and Yan 2003; Zhang, Zhang and Zhang 2007], and multisource data mining in [Wu and Zhang 2003; Wu, Zhang and Zhang 2005; Zhang 2001; Zhang, Wu and Zhang 2003; Zhang and Zaki 2006]. I will try my best to outline some of them in the following subsections.

**2.1. Data mining for multi-users**

While frequent patterns are well-known useful in applications, we advocated to discovering approximate patterns from big data by sampling [Zhang, Zhang and Webb 2003]. These approximate patterns are almost all like "Pattern A" in Figure 2. However, these approximate patterns are dull when data mining applications need support many different users with different accuracies of results. For example, in stock market, stock data mining should support at least two kinds of users as follows. A short-term investor requires a fast approximate result, whereas a long-term investor requires a far more accurate result.

To meet the multiuser applications, we designed an anytime mining algorithm [Zhang and Zhang 2002]. It first takes a dataset, $D_1$, by randomly sampling from a big dataset and mines $D_1$ to obtain the first set $P_1$ of approximate patterns. In this time point, some users can use $P_1$ to their applications. And then, second dataset, $D_2$, by randomly sampling from the big dataset and mines $D_2$ to obtain the first set $P_2$ of approximate patterns. And $P_2$ is reset by integrating $P_1$ and $P_2$ (ensemble learning). In this time point, some users can use $P_2$ to their applications, where P2 is more informative than $P_1$. At $n^{th}$ time of sampling, $P_n$ is obtained by integrating $P_1$, $P_2$, …, $P_n$. Therefore, in the $n^{th}$ sampling, $P_n$ has fused enough information from the big dataset. Figure 3 illustrates the change of rate of approximation to the frequent patterns in the big dataset.

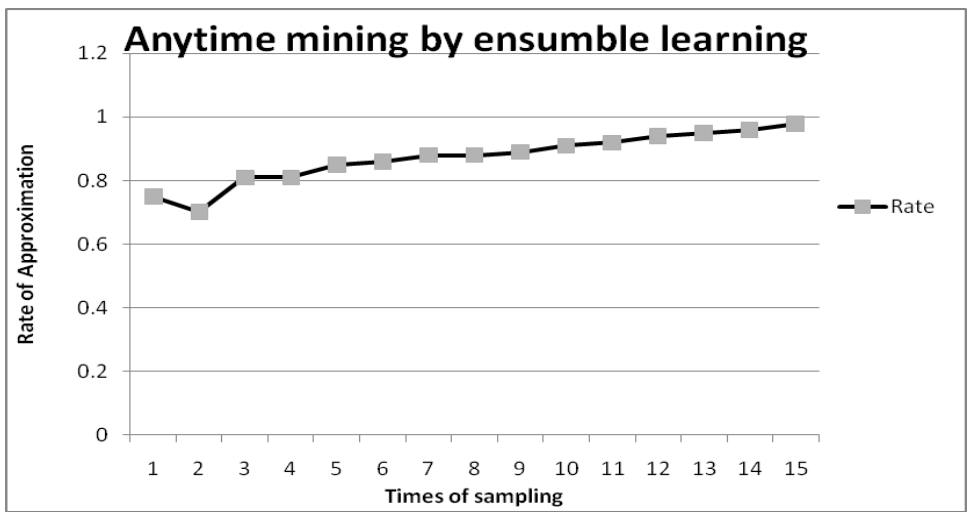

**Figure 3.** Anytime mining

From Figure 3, although there is only 75% rate of approximation to the frequent patterns in the big dataset in the first sampling, they are almost all the frequent patterns with high supports in the big dataset. As we have seen, at the beginning of anytime mining, the rate of the later cannot be better than the former at approximation to the frequent patterns in the big dataset. This is because the support of some frequent patterns in the big dataset is just greater than, or equal to the minsupport. After sampling times are enough large, the rate of approximation will no longer be decreased.

## 2.2. Mining dynamic big data

Mining dynamic data is useful to many practical applications in, such as real-time monitoring and tracking, behaviour prediction, and pattern maintenance. Traditional dynamic data mining is developed against incremental data that cannot discover trend patterns hidden in incremental data. In [Zhang, Zhang and Yan 2003], we designed an algorithm of pattern maintenance for mining incremental dynamic data by weighting. It consists of a new competition mechanism, automation of generating weights for candidate itemsets, and error process. This brings us two benefits to dynamic pattern maintenance. The first benefit is from the competition mechanism that can support the identification of those trend patterns like the "Pattern C" in Figure 2. Another is from the error process that the mining algorithm works on only those incremental data during a period of time. In an applied context, our algorithm is recent-biased and can identify trend patterns, compared with traditional incremental mining algorithms.

From the above, extant dynamic data mining algorithms are only developed against incremental data. These algorithms are difficult to support the pattern maintenance after data delete or data modification. Data delete and modification are two important operations of data update in database management systems. Therefore, we established the decremental mining method that is applied to maintain patterns after acting data delete or data modification to a given dataset in [Zhang, Zhang and Zhang 2007]. From the view of data update, our increment and decrement mining algorithms have formed a complete system that can fully support the pattern maintenance for any data update operations (data delete, modification and append). This system certainly fills in a gap in dynamic data mining.

## 2.3. Mining multisource data

The winner of 2014 ACM Turing Award, Professor Michael Stonebraker pointed out, the big variety of big data means attempting to cope with data arriving from too many data sources, which results in a daunting data integration challenge. Therefore, the efficient mining and utilization of multisource data (MSD) have been one of key issues in interdiscipline of database, artificial intelligence and statistics for decades.

The earliest of MSD mining approach is belonged to the brute-force method, i.e., all data sources are massed together, so as to discover globally useful patterns from the massed dataset. It is often impossible to support such big data discovery because of the capability of both the storage and computation. The modified method is first to select relevant databases. And then, the relevant databases are centralized for pattern discovery. This massed dataset may also be very big and cannot be mined in memory. In particular, this centralized-style mining is inevitable to destroy some useful information, such as the distribution of models, locally burst patterns and exceptional patterns hidden in MSDs.

Different from centralized-style mining algorithms, we designed some efficient strategies for mining MSDs from our data partitioning view, mainly including local pattern analysis [Zhang 2001; Zhang, Wu and Zhang 2003] and data source clustering [Wu, Zhang and Zhang 2005].

**Local pattern analysis**: Local pattern analysis is a fusion procedure as follows. At a node (or a branch), all patterns discovered in its child nodes are taken as local patterns input to the fusion. And then, the data in the node is mined with our data partitioning method if it is applicable, and the identified patterns are also taken as local patterns input to the fusion. Finally, the global patterns at this node are generated by fusing these local patterns. This local pattern analysis can support both globally and locally needs of data mining, regarded as a new mechanism for mining group-behaviour patterns from MSDs. While traditional centralized-style mining approaches are of raw-data sharing, local pattern analysis is of pattern sharing. It means that the input to MSD mining algorithm is essentially different from traditional methods. And local pattern analysis is certainly easy to understand and implement and has high efficiency. This leads to leads to that it is possible to efficiently carry out MSD discovery at almost any nodes (branches), as well as to protect the privacy of original data for companies or branches. In other words, the local pattern analysis has broken through the bottleneck of the storage and computation faced by the centralized-style data mining methods.

With the local pattern analysis, our approach can efficiently approximate the frequent patterns discovered with centralized-style data mining approaches [Wu and Zhang 2003]. In particular, it can identify many interesting patterns useful to decision-makings in different node-levels, such as the distribution of models, locally burst patterns and exceptional patterns [Zhang 2001; Zhang, Zhang and Wu 2004]. It is these patterns that cannot be identified with centralized-style data mining approaches. Indeed, centralized-style data mining approaches can damage some global useful patterns when massing together all relevant datasets. We illustrate this with an example as follows. Let A and B be two tennis players and competing rule be "Best of three sets". If three scores are 6:4, 0:6 and 6:4, respectively, A is certainly the champion or winner according to the game rule. We now take these three games as three databases. Using the centralized-style data mining approaches, it delivers that B is the championship/winner with score 14:12. This result is certainly at odds with the game rule. In other words, the centralized-style data mining methods can destroy the structural information hidden in MSDs. In practical application, the structural information is a kind of the most important information to support multistage decision-making for group cooperation, which is a kind of important patterns in our MSD mining approach.

**Data source clustering**: The centralized-style data mining methods work on relevant datasets selected according to a given mining task. This means, we must select different sets of relevant databases for different mining applications. The selection of data source is referred to application-dependent, or application driven selection. It is certainly very time-consuming and not the best way. In [Wu, Zhang and Zhang 2005], we built a data source clustering mechanism which can overcome the above weakness of traditional MSD mining methods. With data source clustering, we can only mine the relevant classification of data sources based on local pattern analysis. This also leads to further reduce the amount of input to our mining algorithm. In particular, MSD clustering provides an efficient way of data-source management and sharing.

## 3. Data partitioning

From the above discussions, the data partition should benefit the upgrade of all traditional data mining algorithms as much as possible, as well as be efficient for other data analysis and processing applications. Consequently, this research will introduce some methods for big data partitioning based on [Zhang 2016]. Some of commonly-used rules and metrics are outlined as follows.

> **A. Data partitioning based on features**. (A1) Access records: classifying data into frequent access data, rare/never access data, and other data. (A2) Observations: classifying data into known/labelled data, and unknown/unlabelled data. (A3) Class distribution: classifying data into majority-class data, minority-class data, and other data. (A4) Time series: classifying data into latest data, dated data, and other data. (A5) Cases: classifying data into representative data, and other data.

**B. Proportional partitioning**. With the above meta-rules, the partitioning strategies mainly include (B1) top 5% and other 95%; (B2) top *k* that is similar to (B1); and (B3) syllogistic partitioning, referred to golden partitioning in [Zhang 2016].

As is well-known, a data partitioning strategy is a collaboration of the above A and B portioning rules. To make them useful, we illustrate the syllogistic partitioning in an applied context. In a company, some data records in its big dataset are frequently visited or called for some application objectives, whereas some data records in the big dataset have almost never been visited. Consequently, we can generate a syllogistic partitioning based on the visiting frequency as follows.

- The first section, denoted as $D_{5\%}$, is a set of those data records that are with the top 5% visiting frequency in the big dataset.
- The second section, denoted as $D_{30\%}$, is a set of those data records that are with the second-highest 30% visiting frequency in the big dataset.
- The last section, denoted as $D_{65\%}$, is a set of other data records in the big dataset.

We call $D_{5\%}$, $D_{30\%}$ and $D_{65\%}$ as hot data, warm data and cold data, respectively, as showed in Figure 4. Such a syllogistic partitioning forms a dynamic clustering of the big data. And the partitioning will be varied with the changes of time.

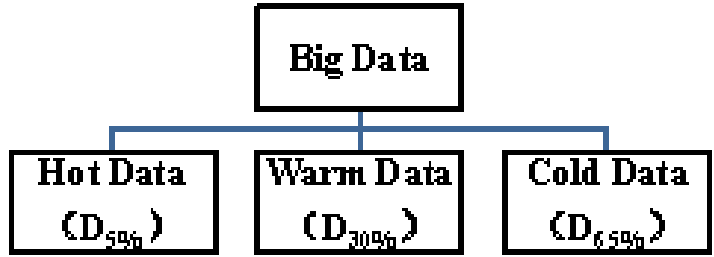

**Figure 4.** Dynamic data partitioning

It must be much more significant to discover patterns and rules from only the hot data, or warm data in the big dataset. If so, we may not mine the cold data in the big dataset. Because $(5\% + 30\%):65\% = 0.539$ is close to the golden ratio 0.618, the syllogistic partitioning was also called as golden partitioning [Zhang 2016]. To make this partitioning actionable, [Zhang 2016] designed a procedure of automatically generating a syllogistic partitioning and maintaining it as follows.

**Procedure** 1. Syllogistic partitioning
**Step 1. Initial** the $D_{5\%}$, $D_{30\%}$ and $D_{65\%}$ by randomly selecting 5%, 30% and 65% data records, respectively, from a big dataset, if we have no knowledge to the big dataset;
**Step 2. Count** the visiting frequency of all data records in the big dataset after the big dataset is used in applications for a certain window of time;
**Step 3. Update** the $D_{5\%}$, $D_{30\%}$ and $D_{65\%}$ by the visiting frequency of the data records in the big dataset;
**Step 4. Output** the $D_{5\%}$, $D_{30\%}$ and $D_{65\%}$;
**Step 5. If** it reaches another window of time, **then** go to Step 2.

Different from the static data partitioning methods, our syllogistic partitioning is dynamically generated in real applications. It well grasps the groupment behaviours of a company to call its big dataset, as well as measures which data are really useful in applications.

## 4. Conclusion

We have illustrated that data partitioning should be an important strategy for mining big data and showed some interesting findings hidden in a big dataset which cannot be discovered by traditional data mining methods. Further, we introduced some data partitioning approaches that can automatically generate data subsets according to the use of the big dataset. This has been collected in some articles concerning big data discovery, such as [Menandas and Joshi 2005] and [Pitre and Kolekar 2014]. Menandas and Joshi thought that our local pattern analysis has laid a foundation for global knowledge discovery in multisource data mining. This theory provides a solution not only for the problem of full search, but also for finding global models that traditional mining methods cannot find. Local pattern analysis of data processing can avoid putting different data sources together to carry out centralized computing.

On the other hand, Pitre and Kolekar pointed out, in case of design of data mining algorithms, knowledge evolution is a common phenomenon in real world systems. But as the problem statement differs, accordingly the knowledge will differ. For example, when we go to the doctor for the treatment, that doctor's treatment program continuously adjusts with the conditions of the patient. Therefore, the local pattern analysis provides a nice solution. This is perhaps the reason why the prestigious journal "*Data Mining and Knowledge Discovery*" provided a special issue to study the local pattern analysis strategy [Zhang and Zaki 2006].

**Acknowledgment**

This work is supported in part by the China "1000-Plan" National Distinguished Professorship; the China 973 Program under grant 2013CB329404; the China Key Research Program (Grant No: 2016YFB1000905); the Natural Science Foundation of China under grant 61672177; the Guangxi "Bagui" Teams for Innovation and Research; the Guangxi Collaborative Innovation Center of Multi-Source Information Integration and Intelligent Processing.